# Flexible MEC service consumption through edge host zoning in 5G networks

Miltiades C. Filippou (*), Dario Sabella (*), Vincenzo Riccobene (**),
(*) Intel Deutschland GmbH; (**) Intel Labs Ireland;
email: {miltiadis.filippou, dario.sabella, vincenzo.m.riccobene}@intel.com

*Abstract* — **Multi-access Edge Computing (MEC) is commonly recognized as a key supporting technology for the emerging 5G systems. When deployed in fully virtualized networks, i.e., following the Network Function Virtualization (NFV) paradigm, it will enable a multitude of new applications and use cases. However, the growing number of devices, combined with the vastly increasing traffic demand, call for low End-to-End (E2E) latency packet transfer and processing in an NFV environment, both in user and control plane. In this paper, focusing on control plane packet traffic, we investigate the general case of a MEC application consuming a MEC service running on a different MEC host. To enable flexible MEC platform service consumption at different localities, based on a state-of-the-art statistical model of the total processing time, we define latency-aware** *proximity zones* **around MEC servers hosting MEC application instances. Exemplary scenarios exhibit the E2E performance benefit of introducing the awareness of proximity zones around MEC hosts and service producing MEC application instances. This performance-aware service consumption will be beneficial in view of the future evolution towards distributed computing systems.**

*Keywords* —*5G, Multi-access Edge Computing, Network Function Virtualization, service consumption, latency*

## I. INTRODUCTION

New communication systems will need to satisfy an increasing need for data traffic communication, by connecting a growing number of devices and heterogeneous traffic flows, with more stringent Quality-of-Service (QoS) requirements [1]-[3]. These challenges are, in their turn, translated into a plurality of demands on the capabilities of new communication and *data processing* systems, starting with early 5G deployments and expanding beyond current systems, since the trend is not going to stop through the years.

To overcome these issues, not only wireless network improvements are needed, but also enhancements of computation systems for packet processing. *Edge computing* is commonly recognized as a key ingredient of future 5G systems [4] [5] that will provide operators with the flexibility and reconfigurability required to satisfy the ever increasing traffic demand and the related tight QoS requirements. More concretely, (i) low latency communication due to the proximity to end users and (ii) the availability of real-time network information, such as radio conditions and network statistics, are important advantages of this technology, which is standardized by the European Telecommunications Standards Institute (ETSI) and known as Multi-access Edge Computing (MEC). Fig. 1 shows an exemplary deployment of a 5G communications system, overlaid by MEC infrastructure, along with instantiated MEC applications and MEC services.

### A. MEC technology overview & APIs

MEC is the international standard introduced by the ETSI Industry Specification Group (ISG) for edge computing [6] [7]. MEC offers to application developers and content providers cloud-computing capabilities and an Information Technology (IT) service environment at the edge of the network. It provides an environment in proximity to the end-user, ultra-low latency, high bandwidth, and access to added-value information through MEC Application Programming Interfaces (APIs), e.g., real-time access to access network, context information, location awareness, etc. MEC will permit to open the cloud infrastructure to operators, service providers and third parties (i.e., application developers and content providers), helping to meet the demanding QoS requirements of new 5G systems. Due to its access-agnostic nature, MEC guarantees also a smoother deployment independent of the underlying Radio Access Network (RAN) (e.g., Long Term Evolution (LTE), 5G New Radio (NR), or other non-3GPP radio access).

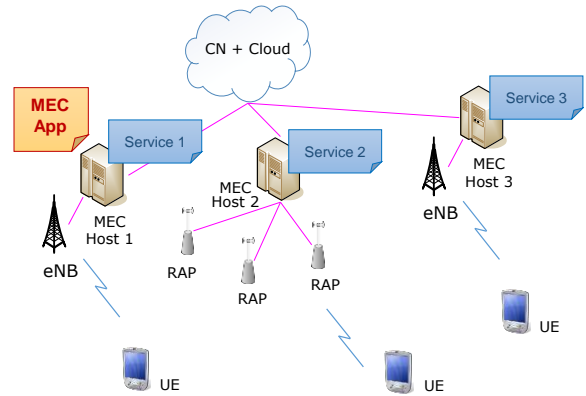

Fig. 1: Example of a a 5G communication system with MEC hosts deployed across all the territory. A MEC application and three MEC services are instantiated at different MEC hosts.

MEC deployments, as well as all communication network elements, will be mainly based on virtualized infrastructure, following the Network Function Virtualization (NFV) paradigm [8]. Nevertheless, the growing number of devices, together with the continuously increasing traffic demand, will create the need for high throughput/ low delay packet transfer and processing in an NFV environment. For this reason, *E2E latency* is a key QoS metric for MEC, and data traffic performance, both in the user plane and the control plane. In this paper, focusing on MEC service consumption performance (thus, concentrating on control plane packets), we analyze the general case of a MEC application consuming a MEC service, through ETSI MEC APIs running on different MEC hosts. In fact, ETSI ISG MEC specifies a set of RESTful APIs as standardized interfaces, which can be used by application developers, in order to access Radio Network Information (RNI API [9]), location information (Location API [10]), and other types of data pre-processed either by the MEC platform or by instantiated MEC applications. In particular, thanks to the RNI API, context information from the RAN can be provided to user level applications or other services for network performance and QoS improvements. As a further example, ETSI ISG MEC is introducing a "MEC V2X API" to assist the MEC system in exposing a set of information to applications, which permit developers, car Original Equipment Manufacturers (OEMs) and their suppliers to implement Intelligent Transportation System (ITS) services in an interoperable way, across different access networks, owned and managed by different Mobile Network Operators (MNOs) and vendors.

*B. MEC service consumption – our contributions*

Since service consumption performance is critical for the access of MEC API information by an instantiated MEC application, in this paper, we propose a MEC QoS awareness enhancement, as a possible contribution to the ETSI MEC standard. This enhancement consists in enabling flexible MEC platform service consumption (locally, or in remote MEC hosts of the same MEC system, or across different MEC systems), by defining QoS/cost-aware *proximity zones* around MEC servers and service-producing MEC application instances. In particular, inspired by the current developments in the ETSI MEC standard, along with the development of various latency-critical applications, we focus on the general case of a MEC application seeking to consume a MEC service, possibly running on a different locality (e.g., MEC host), and highlight the importance of a low-latency connection between the producer and the consumer of that service. In further detail, our contributions are the following:

- We propose a method to define *latency-aware proximity zones* for MEC servers hosting service-producing MEC application instances to enable a flexible use of MEC platform services consumption. This method relies on (a) a statistical model of the total processing (equivalently, service consumption) time at a host and/or (b) generic gathering of proximity measurements from neighboring MEC hosts and/or MEC systems, performed by the MEC Orchestrator (MEO) [11]. In particular, since we consider fully virtualized 5G networks, control packet traffic for MEC service consumption is modeled assuming a NFV-based MEC deployment, and the corresponding latency refers to the implementation of the MEC system in a NFV infrastructure.

- To accomplish a minimum latency MEC service consumption by an instantiated MEC application, the defined proximity zones are taken into consideration by means of a *signaling protocol* among the involved MEC entities. As a result of such signaling, the consumption of the MEC service can be either endorsed by the MEO, or a MEC application instant migration event may need to be triggered to prevent both service consumption failures and non-tolerable delays.

- We complement the study by providing examples of MEC service consumption based on the availability of *statistical performance information* available at the system's MEO.

The remainder of the paper is organized as follows: in Section II, we describe the reference 5G communication system, overlaid by MEC system infrastructure, as well as a set of quantities relevant to packet traffic and packet processing at the deployed MEC hosts, which will be taken into account for constructing the MEC proximity zones. In Section III, we pose the problem of MEC service consumption by a MEC application, when the service consumer(s) and service producer(s) are instantiated at different MEC entities/ localities, and we introduce the proposed solution. Section IV includes examples illustrating the benefits of our proposed solution, while, Section V concludes the paper.

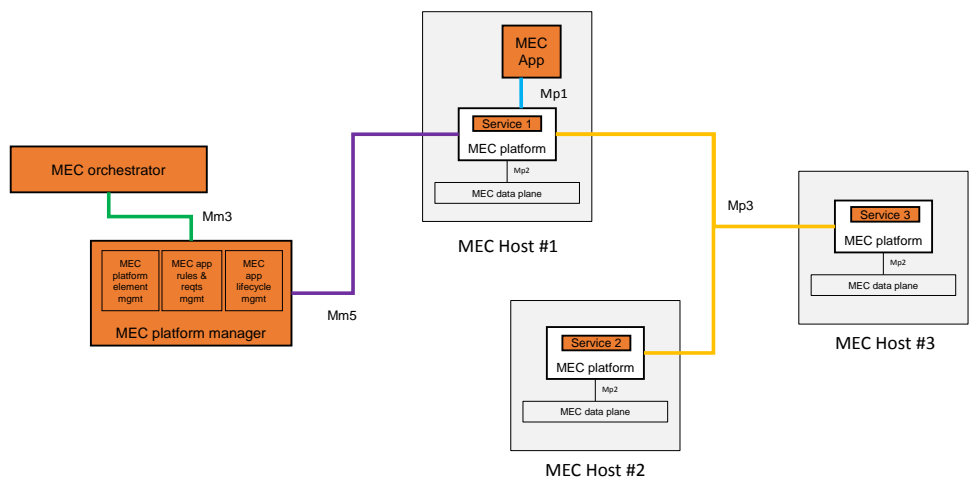

Fig. 2: Exemplary topology of a MEC system consisting of three MEC hosts; a MEC app is instantiated at MEC Host # 1, while, three services are running across the MEC system.

## II. SYSTEM SETUP & CONTROL PLANE PACKET TRAFFIC PERFORMANCE

We consider a 5G communication system with MEC hosts deployed across a geographical area. Although the setup is extendable to a multi-MEC system one, for the sake of simplicity, we consider only one MEC system composed of different MEC hosts, where, each MEC host is associated to, at least one Base Station (BS) (e.g., eNB, 5G NB or Radio Access Point (RAP)) in the network. The most general case corresponds to a MEC application (denoted as "MEC app" in the remainder of the document) running on a MEC host, which needs to consume MEC services instantiated within the (same) MEC system. The queried services are assumed available in the MEC system, however, not necessarily running at the same MEC host (also termed after as "MEC server" in the remainder of the paper).

### A. MEC system setup – functional elements and reference points

Following the ETSI MEC reference architecture [11], Fig. 2 depicts an exemplary topology of a MEC system consisting of three MEC hosts within which MEC platforms run different MEC services, a MEC Platform Manager (MEPM), a MEO, as well as the various interfaces/ interconnections between these entities, i.e., interface Mm3 connecting the MEO with the MEPM, interface Mm5 connecting the MEPM with the MEC platforms, as well as interface Mp3 inter-connecting the MEC platforms of the system, along with interfaces Mp1 and Mp2 within each MEC host. It is assumed that a MEC app is running on MEC Host # 1, which is potentially in need of consuming some of MEC services 1, 2 and 3.

### B. Statistical models for MEC control plane packet queuing & processing

In order to evaluate the performance of control plane packet traffic, in the general case of a MEC app consuming a MEC service, we may need to apply a statistical model of the delay performance of IP packets forwarded via Mp1/Mp3 interfaces, either locally (i.e., within the same MEC host) or involving different MEC hosts. Since the assumption is to run MEC in a NFV environment, a good reference for the present analysis is a statistical model on packet queuing and packet processing for softwarized network functions that was developed and cross-validated in [12]. As per this model, using discrete-time analysis, two packet queues are involved: a peripheral queue and a central (Central Processing Unit – CPU/processing) queue. In [12], the two queue models are combined into an overall model to express the distribution of the total processing time of a given packet. This is composed of (i) the waiting time in the peripheral queue, (ii), the waiting time in the central queue and (iii) the packet service time in the latter queue. It is noted that the packet waiting time is limited by a maximum tolerable value of $L$ time units, meaning that, packets that arrive and would have to wait longer than $L-1$ time units are rejected. Thus, given these assumptions and under stationary conditions, the probability distribution of the total processing time of the $i$-th packet of a given batch, $D_i$, is given by the following

$$d_i(k) = w_i(k) * u(k) * b(k) * \ldots * b(k), \quad (1)$$

$k = 0, 1, \ldots$, where, $w_i(k)$ is the probability distribution of the waiting time of the $i$-th packet in the peripheral queue, $u(k)$ is the distribution of unfinished work in the system before the arrival of the next batch, $b(k)$ is the distribution of packet service time and the convolution of $b(k)$ takes place $i$ consecutive times. It should be noted that operator $(*)$ stands for the discrete convolution operation. As a result, the probability distribution of the random variable representing the total processing time for all packets, $D$, can be determined by the expression that follows

$$d(k) = \sum_{i=1}^{\infty} P(X = i) \, d_i(k) = \sum_{i=1}^{\infty} x(i) \, d_i(k), \quad (2)$$

where, $X$ stands for the random variable expressing the batch size and $x(i)$ denotes the probability distribution of the latter quantity, i.e., $x(i) = P(X = i)$.

Having presented the communication system model and the MEC system setup, in the following section we introduce the proposed design of latency-aware proximity zones for MEC service consumption.

## III. A MEC HOST PROXIMITY ZONING FRAMEWORK FOR LOW-LATENCY MEC SERVICE CONSUMPTION

### A. Designing the MEC proximity zones

Focusing on the system setup of Fig. 2, let us assume that a MEC app instantiated at MEC Host #1 is in need of some of the three MEC services running on MEC Hosts # 1-3, respectively and are accessible from the running MEC app through the Mp1/Mp3 reference points. To evaluate the price to pay in terms of delay, so as for the MEC app to consume a specific MEC service, the proximities of MEC Hosts # 2 and 3 need to be measured, having MEC Host # 1 as a reference (see Fig. 3). Then, based on a proximity criterion, the MEC hosts can be grouped into performance zones accordingly and the zone data (i.e., zone ID and the IDs of the corresponding MEC hosts) can be stored in the MEC system (MEO) in an asynchronous fashion.

*Table I: Classifying MEC hosts into proximity zones according to a utility criterion (MEC Host # 1 is the reference)*

| Proximity zones | Proximity zone characteristics | | |
|---|---|---|---|
| | *Minimum cost units* | *Maximum cost units* | *MEC hosts of the zone* |
| 1 | 0 | 5 | 1 |
| 2 | 0 | 10 | 1, 2, 3 |

To accomplish that procedure, the MEO will form a data structure (e.g., a table) defining zones of MEC hosts, based on the accumulated latency needed to reach the reference MEC host running the MEC app (i.e., MEC Host # 1 in our example), either via measurements (*online zoning*), or, by exploiting a statistical model, as the one presented in Section II.B (*offline zoning*). In fact, the MEO is the entity responsible for gathering, classifying and storing proximity measurements, as it has an overall view of the MEC system topology, the available resources and the available MEC services [11]. Table I, accompanied by Fig. 3, provide an example of such proximity-based classification maintained at the MEO. It should be noted that the cost values in this example are just indicative, as well as that a proximity zone which is an enclave of another proximity zone is also part of the latter one.

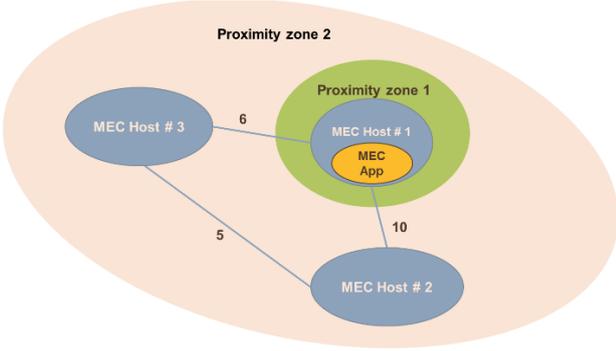

Fig. 3: Visualization of MEC host proximity zones with reference to the running MEC app, according to the utility-based classification of Table I.

As shown in Fig. 3, only MEC Host # 1 belongs to proximity zone 1, whereas, proximity zone 2 incorporates MEC hosts, the hosted MEC services in which can be reached, at a higher cost, or, equivalently, worse MEC app-to-service latency performance. It should be noted that the construction of the MEC proximity zones should be updated each time the MEC system topology is altered, for example, when more MEC hosts are deployed over a given geographical area, and/or, when the delay measures are updated (e.g., because the physical interfaces inter-connecting the MEC servers are upgraded). Also, to verify the validity and timeliness of the statistically constructed proximity zones, in case a statistical latency model is utilized, real measurements can be used by the MEO, for instance, to update the parameters of the statistical model (e.g., the distribution of the delay between two MEC hosts at which a service producer and a service consumer are instantiated).

Having defined MEC proximity zones according to a performance based criterion, in the next section, we will introduce a signaling protocol among the various MEC system entities with the aim of achieving QoS-compliant service consumption by a given MEC app instantiated at a host of the focused MEC system.

### B. A zone-aware signaling protocol for delay-efficient MEC service consumption by a MEC app

In the previous section we have described the design of MEC proximity zones, which supports the MEO to evaluate which services can be consumed (or not) by a certain MEC app. According to this information, the system may decide that a certain remote service cannot be consumed because the resulting service consumption latency is too high. As an alternative, and to avoid MEC app failures, the MEO may recommend the *migration* of the MEC app instance to another MEC host, closer to the needed service (of course, in case the user plane E2E delay, i.e., between the UE and the MEC app instant is still kept at a reasonable level). This case of MEC app migration is depicted in Fig. 4. According to this example, in the beginning, only MEC app 1 running on MEC Host # 3 directly consumes MEC service 1 instantiated at MEC Host # 4, as it is within its proximity zone. However, when MEC app 2 running on MEC Host # 1 requests to consume MEC service 1, since the service is outside its proximity zone, the MEC system will need to evaluate whether a MEC app instance migration to, e.g., MEC Host # 2, would satisfy the performance requirements for consuming the service, also accounting for the latency experienced by the UE once the MEC app instance migration is performed. If the E2E requirement is satisfied, the MEC app instance can be relocated to MEC Host # 2, therefore, the needed service will be within its proximity zone, which will allow for low-delay service consumption.

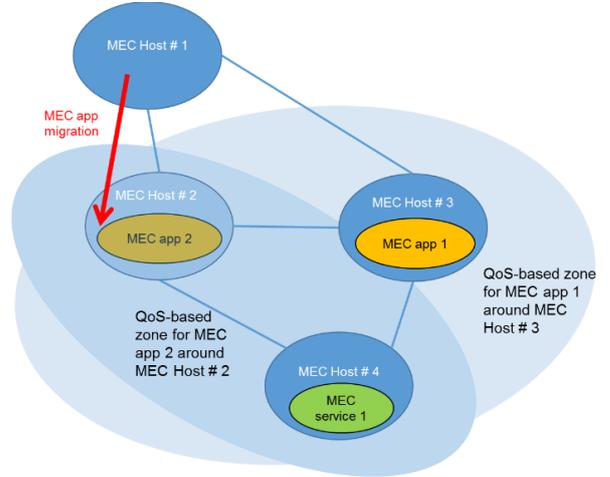

Fig. 4: Example of MEC app migration, for latency-reduced MEC service consumption.

Focusing on the example depicted in Fig. 4 and concentrating on MEC app 2, in the following message sequence chart (Fig. 5), a signaling protocol among the involved MEC entities, triggered by the service-seeking MEC app, is explained to support MEC service consumption based on the above described QoS-aware zoning criterion.

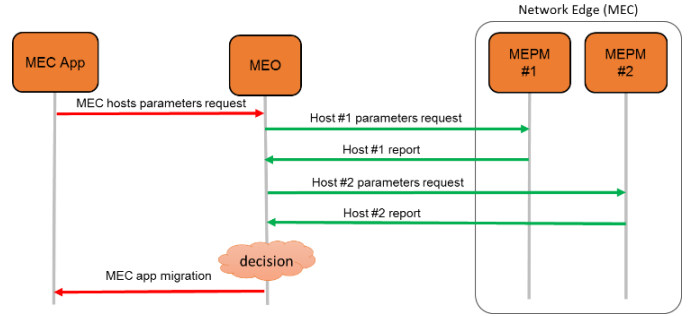

Fig. 5: Signaling protocol among MEC functional entities to support proximity zone-based low-latency MEC service consumption.

The internal decision taken by the MEO (which is not to be specified in a standard) is based on the information obtained from the different MEC hosts in the system, and it aims to facilitate a low-latency MEC service consumption by MEC app 2, taking into account the defined proximity zones. According to this specific example, the last message of the chart is a MEC app migration indication, while it may also happen that no migration is needed, i.e., the MEC app can readily consume the queried service. In the latter case, the MEO answer is empty, i.e., a HTTP "200=OK" status response.

*Note*: this scenario can be applicable also in the case of consuming MEC services running in different MEC systems (e.g., associated to different network operators). In this case, the principle of *MEC-as-a-service* is applied to the communication among all involved entities: considering a two-system example, MEC system # 1 is exposing information through its MEO-1, and a MEC app running in

MEC system # 1 is able to consume services instantiated in another MEC system # 2, which is exposing information to the MEO-1 through its MEO-2. After information exposure, MEO-1 will take the final decision, based on the overall acquired knowledge of proximity zones across MEC systems 1 and 2. Finally, the MEC app running in MEC system # 1 will receive a final decision/recommendation response from its MEO-1, as already depicted in Fig. 5.

## IV. EXEMPLARY CASE STUDIES - THE PRACTICAL PERSPECTIVE

### A. MEC service consumption based on statistical packet latency information at the MEO

Let us assume a MEC system deployment, composed of three MEC hosts, such as the one of Fig. 2, where, a MEC app is instantiated at MEC Host # 1 and MEC services are instantiated at MEC hosts # 1, 2 and 3, respectively. Assuming that, for the MEC app, it is enough to consume any of these three MEC services, the MEO, as described before, will be tasked to recommend a decision upon the selection of the service producer (or, information source).

For our case study, motivated by [12] [13], we assume that, both locally, as well as, for each MEC host-to-host communication, $i$, the packet inter-arrival time follows a distinct random distribution, with a distinct expected value of $\mu_A^i$ μsec, while, the packet processing time at the CPU follows another random distribution with an expected value of $\mu_B$ μsec. The normalized arrival rate, which is the main contributor to system load, is defined as $\alpha_i = \frac{\mu_B}{\mu_A^i}$, whereas, for the inner (CPU) queue, we assume that a GI/GI/1-L system is implemented.

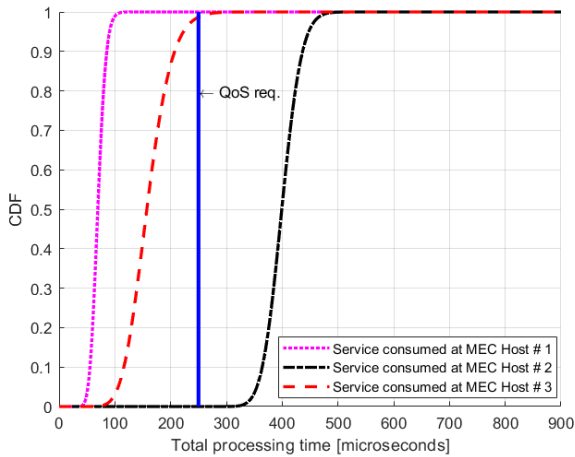

Fig. 6: MEC service consumption based on statistical information on the total processing time, $D_i$.

Assuming that the abovementioned statistical quantities (i.e., the random distributions and their parameter values) can be inferred by the MEO via processing large amounts of measurements, or, can be directly known via adopting the model in [12], the MEO will be able to construct Cumulative Distribution Function (CDF) curves of the total processing time for every service consumption option, $D_i, i = 1,2,3$. As evident by Fig. 6, given a QoS requirement of e.g., 250 μsec with a confidence level above 95%, the MEC app will consume the MEC service instantiated at MEC Host # 3, if the sought after service is not already running at the same server (i.e., at MEC Host # 1). For comparison, in the absence of any proximity zoning, where, the MEC app would *randomly* choose an available service to consume, the 95[th] percentile of the resulting total processing time would be even doubled (i.e., in the case, where the MEC service running at MEC Host # 2 would be chosen over the MEC service running at MEC Host # 3).

### B. Packet transfer performance based on COTS products

In the scope of this work, the main Key Performance Indicator (KPI) of interest is the latency, which also represents the most positively impacted KPI by the application of the MEC concept to telecommunication networks. According to [14], the main E2E latency components in communication systems are the *propagation delay*, and the *processing delay*.

In terms of propagation delay (e.g., data transmission, serialization, data protocols, routing and switching, as well as queuing), the data transmission is the most impactful element of the chain. For instance, using fiber optics cable connections, the maximum speed reachable is ~4.77μs/m [14].

With regards to the processing delay, the application of NVF-based MEC can lead to interesting delays due to the fact that the network packet processing happens inside a virtual machine. This results in a relatively deterministic delay in case of point-to-point connections, as well as in the case of connections including predetermined paths or a very small number of network devices between a source and a destination. The MEC scenario falls into this set of cases, therefore, we can assume that the latency between MEC hosts generally depends on the geographical distance between them and can be either accurately estimated or measured by monitoring systems.

In addition to the above, the E2E delay will also be differently impacted by different Virtualized Network Functions (VNFs), in relation to the type of processing required by the application/service. In [15], some test results for a simple VNF (a router simply forwarding packets) are presented, demonstrating that the delay for the processing of packets can be less than 100μs. These results have been obtained using open source technologies like Linux Operating System (OS) and Open vSwitch (OvS). Authors of [15] also demonstrate that the processing delay can be reduced significantly (more than 50%) when using Real Time OS kernels and the DPDK OvS [16]. Another important result demonstrated is that the deployment of a service chain composed of two VNFs of the same kind on the same physical host leads to a latency increase of ~1.5 times, as compared to the single VNF case.

*TABLE II: One-way latency measurements from [16] for the OvS and DPDK setup.*

| Packet size (bytes) | Latency 1 VNF OvS (μs) | Latency 2 VNFs OvS (μs) | Latency 1 VNF DPDK (μs) | Latency 2 VNFs DPDK (μs) |
|---|---|---|---|---|
| 64 | 75 | 128 | 28.4 | 44.74 |
| 256 | 132 | 202 | 80.92 | 132.5 |
| 1500 | 181 | 268 | 88.34 | 170.23 |
| **Average** | **129** | **199** | **66** | **116** |

Authors in [17] report, as a result of a complete characterization of DPDK OvS, a linearly increasing median from 7.8μs (99th percentile: 8.2μs) to 13.9μs (99th percentile: 23.0μs) for the latency from pNIC (physical Network Interface Controller (NIC)) to pNIC through a DPDK OvS, using different levels of network stress.

Analyzing the aforementioned references, it is possible to estimate the expected E2E latency taking into account the cable length between two MEC instances and the placement of the services/applications. The numbers reported in Fig. 7 are generated averaging numbers corresponding to experiments with different packet sizes and network workload conditions from the references discussed above and for this reason, it is useful only to treat them as general indications (i.e., in terms of order of magnitude) of the latency of each single component of the system. Of course, the measured latency will be different from these numbers, in a case-by-case manner, depending on the size and the amount of packets to be processed.

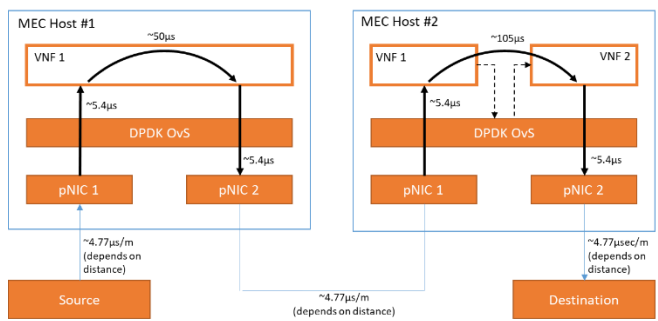

Fig. 7: Indication of the order of magnitude of latency of different system components when using DPDK OvS.

## V. CONCLUSION

In this paper, we have studied the performance of MEC systems, when deployed in NFV environments. In particular, we have analyzed the general case when a MEC app needs to consume local or remote MEC services (i.e., running in different hosts) and the related delay issues. To enable a flexible use of MEC platform services consumption at different localities, we proposed the design of latency-aware proximity zones around MEC servers hosting service-producing MEC app instances. Then, we analyzed the related packet traffic in a NFV-based MEC deployment example, and exploited a state-of-the-art statistical model on the total processing time for drawing decisions on MEC service consumption scheduling. Examples illustrated the E2E performance potential of introducing the awareness of proximity zones. Looking forward, such latency-aware service consumption will be beneficial, especially when considering emerging distributed computing paradigms, where a growing number of heterogeneous connected devices, producing and consuming MEC services, will further emphasize the need for optimized data traffic communication. Future work will potentially include relevant contribution to the ETSI MEC standard, alignment with 3GPP specifications, and possibly, extensive numerical evaluations, e.g., comparing the latency performance of MEC system deployments of different densities and processing capabilities.